\documentclass{eptcs}
\providecommand{\event}{ACL2 2020} 
\usepackage{underscore}           
\usepackage{alltt,fancyvrb,graphicx,latexsym,pifont,amsmath,amstext,amssymb,amsfonts,epic,eepic}

\begin{document}

\numberwithin{equation}{section}

\pagenumbering{arabic}

\title{Formal Verification of Arithmetic RTL:\\ Translating Verilog to C++ to ACL2}
\author{David M. Russinoff
\email{david@russinoff.com}}
\def\titlerunning{Translating Verilog to C++ to ACL2}
\def\authorrunning{D.M. Russinoff}

\maketitle

\begin{abstract}
We present a methodology for formal verification of arithmetic RTL designs that combines sequential logic equivalence checking with interactive
theorem proving.  An intermediate model of a Verilog module is hand-coded in {\it Restricted Algorithmic C (RAC)}, a primitive subset of C
augmented by the integer and fixed-point register class templates of Algorithmic C.  The model is designed to be as abstract and compact as
possible, but sufficiently faithful to the RTL to allow efficient equivalence checking with a commercial tool.  It is then automatically translated
to the logic of ACL2, enabling a mechanically checked proof of correctness with respect to a formal architectural specification.  In this paper,
we describe the RAC language, the translation process, and some techniques that facilitate formal analysis of the resulting ACL2 code. 
\end{abstract}

\section{Introduction}

A prerequisite for applying interactive theorem proving to arithmetic circuit verification is a reliable means of converting an RTL design to
a semantically equivalent representation in a formal logic.  Within the ACL2 community, this has been achieved in at least two industrial settings
by mechanical translation from Verilog directly to the ACL2 logic.  At Advanced Micro Devices, our approach was to generate a large clique of
mutually recursive executable ACL2 functions in correspondence with the signals of a Verilog module~\cite{amd}, a scheme commonly known as
``shallow embedding''. The Centaur Technology verification team has followed a different path~\cite{centaur}, converting an RTL design to a netlist of
S-expressions to be executed by an interpreter coded in ACL2, i.e., a ``deep embedding'' of the design.  Each approach has its
advantages~\cite{boulton}---in this case, the result of the first is more suitable for traditional theorem proving while the second better enables
verification by ``bit-blasting'', but both burden the user with an unwieldy body of ACL2 code, at least comparable in size to the Verilog source.

Theorem proving in general is uncommon in the chip design industry.  A more popular formal verification technique is sequential logic
equivalence checking, by which a design is checked against a trusted model, either a high-level C++ program or a legacy Verilog design,
with the use of a commercial tool such as SLEC~\cite{slec} or Hector~\cite{hector}.  This method has the advantage of being largely automatic and
requiring less expertise of the user, but it suffers from two main deficiencies.  First, the so-called ``golden model'' is usually
trusted solely on the basis of testing, not having been formally verified itself.  The second is the complexity limitations of these tools, which are
likely to render them ineffective unless either the RTL is relatively simple or its design closely matches that of the model.

Here we describe a hybrid solution, initially developed at Intel~\cite{oleary} and now in regular use at Arm~\cite{book}, that combines equivalence
checking with theorem proving in a two-step process.  First, an intermediate model is derived from the RTL by hand, coded in {\it Restricted Algorithmic
C (RAC)}, a primitive subset of C augmented by the register class templates of Algorithmic C~\cite{ac}, which essentially provide the bit manipulation
features of Verilog.  The objective is a high-level model that is more manageable than the RTL but mirrors its microarchitecture
to the extent required to allow efficient equivalence checking with SLEC~\cite{slec}.  This program is then processed by a RAC-ACL2 translator,
which itself involves two steps: a special-purpose Flex/Bison parser~\cite{flex} generates a set of S-expressions, which are converted to ACL2 functions
by a code generator written in ACL2.  Finally, the prover is used to check a proof of correctness of the design with respect to a formal architectural
specification.  Such proofs are supported by an ACL2 library of lemmas pertaining to RTL and computer arithmetic, found in the directory
{\tt books/rtl} of the ACL2 repository.  This directory also contains specifications of the elementary arithmetic instructions of the x86 and Arm
architectures.  The library and the underlying arithmetic theory are documented in~\cite{book}.  The RAC translator resides in {\tt books/projects/rac}.

The primary advantage of this approach over direct translation is that it provides an abstract and readable representation of the design that is
amenable to mathematical analysis, and consequently a compact ACL2 model that is more susceptible to formal proof.  The translation of Arm floating-point
units has been found to result in a reduction in code size by approximately 85\%.  The intermediate model serves other purposes as well, including
documentation, design guidance (in some cases the RAC model has been written before the RTL), and simulation in a C++ environment.  One disadvantage
is the number of software systems involved, each of which may be viewed as a possible source of error.  Another is the significant effort required in
extracting the model from the design.  On the other hand, much of this effort would still be needed for the proof effort.

This methodology has been successfully applied in the verification of a wide range of arithmetic components of Intel and Arm CPUs and GPUs, including
high-precision multipliers, adders, and fused multiply-add modules; 64-bit integer multipliers and dividers; a variety of SRT division and square root
modules of radices 2, 4, and 8; and a radix-1024 divider with selection by rounding.  Examples of these applications, including RAC models, their
ACL2 translations, and proof scripts, may be found in {\tt books/projects/arm}.  RAC is also the basis of the hardware/software co-assurance
work of Hardin.~\cite{dh}

In Section~\ref{rac}, we present the basic features of the RAC language.  The parser and code generator are described in Sections~\ref{parser}
and~\ref{codegen}.  Some established techniques for proving theorems about RAC programs are discussed in Section~\ref{proof}.

\section{Restricted Algorithmic C}\label{rac}

The RAC language consists of a small set of constructs that have been found to be suitable for modeling RTL designs, especially floating-point
units:

\begin{itemize}

\item Basic numerical data types: {\tt bool}, {\tt uint}, {\tt int}, and {\tt enum}s (but no pointers);

\item Composite types: arrays and {\tt struct}s;

\item Simple statements: variable and constant declarations, assignments, type declarations, and assertions;

\item Control statements (with restrictions): {\tt if}, {\tt for}, {\tt switch}, and {\tt return};

\item Functions (with value parameters only);

\item Standard C++ library templates: {\tt array} and {\tt tuple};

\item Algorithmic C class templates: arbitrary width signed and unsigned integer and fixed-point registers, bit manipulation primitives, and logical
operations.

\end{itemize}
An object of a register type is a bit vector of a width specified by a template parameter.  The interpretation of the vector upon evaluation depends
on its type.  An unsigned integer register of width $n$ is evaluated simply as an integer in the range $[0, 2^n)$. For a signed integer of the same
width, bit $n-1$ is interpreted as a sign bit and the value is in the range $[-2^{n-1}, 2^{n-1})$.  Fixed-point registers are similarly interpreted,
except that an additional template parameter indicates the location of an implicit binary point.  For example, an unsigned fixed-point register of
width $n$ with $m$ integer bits holds an $n$-bit vector $x$ with interpreted value $2^{m-n}x$.

The use of integer registers is illustrated in the first example of Figure~\ref{add8}.  By convention, we denote an unsigned (resp., signed) integer
register class of width $n$ as {\tt ui}$n$ (resp., {\tt si}$n$).  In this case, the definition of {\tt add8} has been preceded by the type declarations
\begin{verbatim}
     typedef ac_int<32, false> ui32;
     typedef ac_int<8, true> si8;
     typedef ac_int<9, true> si9;
\end{verbatim}
(This allows us subsequently to avoid some of the cumbersome syntax of C++.)  Thus, the function takes 2 32-bit arguments, from which it extracts
corresponding 8-bit slices, adds them as signed integers, ``saturates'' the sums to lie in the range $[-128, 128)$, and stores them in a
32-bit result vector.  The bit slice extraction method {\tt slc} has a template parameter indicating the width of the slice and an argument
representing the base index.  The {\tt set_slc} method, which writes a value to a slice, takes two arguments, which determine the base index of the
slice and the value to be written, the type of which determines the width of the slice.  In the application of these methods, no disctinction between
signed and unsigned or integer and fixed-point types is recognized.  The same is true of logical operations on registers.  But when a register is
evaluated, the type is relevant.  Thus, in the declaration
\begin{verbatim}
        si9 sumSgnd = aSgnd + bSgnd;
\end{verbatim}
the signed values of {\tt aSgnd} and {\tt bSgnd} are added using unbounded arithmetic (matching the semantics of ACL2) and the result is written
to the signed register {\tt sumSgnd}, truncating if necessary (although here the width 9 is chosen to avoid any loss of data).  Similarly, in the next
line, it is the signed value of that register that is compared to -128.

\begin{figure}
\begin{footnotesize}
\begin{verbatim}               
                                ui32 add8(ui32 a, ui32 b) {        
                                  ui32 result; ui8 sum;            
                                  for (uint i=0; i<4; i++) {       
                                    si8 aSgnd = a.slc<8>(8 * i);   
                                    si8 bSgnd = b.slc<8>(8 * i);   
                                    si9 sumSgnd = aSgnd + bSgnd;   
                                    if (sumSgnd < -128)            
                                      sum = -128;                  
                                    else if (sum >= 128)           
                                      sum = 127;                   
                                    else                           
                                      sum = sumSgnd;               
                                    result.set_slc(8 * i, sum);}
                                  return result;}                    
\end{verbatim}
\end{footnotesize}
\caption{A signed integer adder}\label{add8}
\end{figure}

The absence of pointers and reference parameters from our language implies that a function simply returns a value without side-effects and that the
C parameter-passing mechanism for arrays is disallowed.  The purpose of including the two standard library templates is to compensate for these
restrictions: the {\tt array} template allows arrays to be passed by value and {\tt tuple} provides the effect of multiple-valued functions (and in
fact its use is restricted to this context).

A variety of control restrictions are imposed to facilitate translation to ACL2.  Note that {\tt do}, {\tt while}, {\tt continue}, and {\tt break}
(except in a limited way within a {\tt switch} statement) are excluded.  A {\tt for} loop, in order to ensure admissibility of the generated recursive
function (see Section~\ref{codegen}), is required to have the form

\begin{alltt}
  for ({\it init}; {\it test}; {\it update}) \{ ... \}
\end{alltt}
where

\begin{itemize}

\item {\it init} is either a declaration of, or an assignment to, the loop variable {\it var}, which must be of type {\tt uint} or {\tt int}.

\item {\it test} is either a comparison between the loop variable and a numerical expression
of the form {\it var} {\it op} {\it limit}, where {\it op} is \verb!<!, \verb!<=!, \verb!>!,
or \verb!>=!, or a conjunction of the form $\mathit{test}_1$ \verb!&&! $\mathit{test}_2$, where $\mathit{test}_1$
is such a comparison and $\mathit{test}_2$ is any boolean-valued term.

\item {\it update} is an assignment to the loop variable.

\end{itemize}
The combination of {\it test} and {\it update} must
guarantee termination of the loop.  The translator derives a {\tt :measure} declaration from {\it test}, which is
used to establish the admissibility of the resulting function.
In some cases, the {\it test} may be used to achieve the functionality of {\tt break}. For example, instead of

\begin{alltt}
  for (uint i=0; i<N; i++) \{if ({\it expr}) break; ... \}
\end{alltt}
we may write
\begin{alltt}
  for (uint i=0; i<N && !{\it expr}; i++) \{ ... \}
\end{alltt}
This feature may also be used in cases where the equivalence checker is unable to establish an absolute upper bound on the number
of iterations executed by the loop, which is required for the ``unrolling'' that is performed by the tool.  Thus, while ACL2 has no trouble
establishing termination
of either of the above loops, SLEC may require something like the following, which may be used when $N$ is known never to exceed 128:
\begin{alltt}
  for (uint i=0; i<N && i<128; i++) \{ ... \}
\end{alltt}
Further restrictions are imposed on the placement of {\tt return} statements.  We require every function body to be a statement block that
recursively satisfies the following conditions:

\begin{itemize}
\item [(1)] The statement block consists of a non-empty sequence of statements;
\item [(2)] None of these statements except the final one contains a {\tt return} statement;
\item [(3)] The final statement of the block is either a {\tt return} statement or an {\tt if}$\dots${\tt else} statement of which
each branch is either a {\tt return} statement, an {\tt if}$\dots${\tt else} statement that satisfies this condition, or a statement block
that satisfies all three of these conditions.
\end{itemize}

The design of a program in this language is generally a compromise between two opposing objectives.  On the one hand, a higher-level model
is more susceptible to mathematical analysis and allows a simpler correctness proof.  On the other hand, successful
equivalence checking of a complex design generally requires a significant amount of proof decomposition, using techniques
that depend on structural similarities between the model and the design.

As a rule of thumb, the model should be as abstract as possible while performing the same essential computations as the design.  For example,
in the case of a Booth multiplier, it is advisable to replicate the partial products and each level of the compression tree in order to allow
the necessary decomposition.  For a high-precision SRT divider, we find that successful equivalence checking requires a bitwise match between the
partial remainders and quotients on each iteration.

\begin{figure}
\begin{footnotesize}
\begin{verbatim}          
                                ui6 CLZ64(ui64 x) {
                                  assert(x != 0);
                                  bool z[64];
                                  ui6 c[64];
                                  for (uint i=0; i<64; i++) {
                                    z[i] = !x[i];
                                    c[i] = 0;}
                                  uint n = 64;
                                  for (uint k=0; k<6; k++) {
                                    n = n/2; // n = 2^(5-k)
                                    for (uint i=0; i<n; i++) {
                                      c[i] = z[2*i+1] ? c[2*i] : c[2*i+1];
                                      c[i][k] = z[2*i+1];
                                      z[i] = z[2*i+1] && z[2*i];}}
                                  return c[0];}

\end{verbatim}
\end{footnotesize}
\caption{A leading zero counter}\label{clz}
\end{figure}

Abstraction and simplicity are achieved by eliminating implementation details and ignoring timing and cycle structure, all of which may be done
without adversely affecting the equivalence check.  Settling on the proper level of abstraction often requires some experimentation.  Consider the
leading zero counter of Figure~\ref{clz}, which executes in logarithmic time with respect to the argument width.  After $k$ iterations of the loop,
where $0 \leq k \leq 6$, the value of the variable $n$ is $2^{6-k}$ and the argument is conceptually partitioned into $n$ slices of width $2^k$.
Each of the low order $n$ entries of the boolean array $z$ is set if and only if the corresponding slice is 0, and when this is not the case, the
corresponding entry of the array $c$ holds the number of leading zeroes of the slice.  This function, which was designed to match the behavior of a
component of a floating-point adder, consists of considerably less code than the Verilog version but allows a fast equivalence check.  It is natural
to ask whether the check would succeed if the function were to be based on a more transparent linear-time algorithm:

\begin{verbatim}
     ui6 CLZ64(ui64 x) {
       int i;
       for (i=63; i>=0 && !x[i]; i--) {}
       return i;}
\end{verbatim}
It does indeed, but the overall SLEC run-time for the adder that includes the function increases from 2 minutes to 22 minutes.

\section{The RAC Parser}\label{parser}

A by-product of the RAC parser is a more readable pseudocode version of a function, with C++ templates, methods, etc., replaced with a syntax that is
more familiar to Verilog programmers.  For example, the {\tt slc} and {\tt set_slc} calls in {\tt add8} are replaced by
\begin{verbatim}
    si8 aSgnd = a[8*i+7:8*i];
    si8 bSgnd = b[8*i+7:8*i];
\end{verbatim}
and
\begin{verbatim}
    result[8*i+7:8*i] = sum;
\end{verbatim}
Its primary purpose, however, is the conversion of a function definition to an S-expression that preserves its structure.
The parser output for the function {\tt add8} is displayed in Figure~\ref{ast}.
Note that some RAC primitives correspond to built-in ACL2 functions, while others correspond to functions defined in the RTL book
{\tt lib/rac}: {\tt BITS} and {\tt SETBITS} extract and assign slices of a bit vector; {\tt LOG<}, {\tt LOG>=}, etc., are comparators that return
1 or 0 (emulating C); {\tt SI} computes the signed integer value of a register of a given width.  Other symbols that appear in the
output---{\tt FUNCDEF}, {\tt BLOCK}, {\tt FOR}, etc.---are not defined as ACL2 functions but are processed later by the code generator.

\begin{figure}
\begin{footnotesize}
\begin{verbatim}
    (FUNCDEF ADD8 (A B)
      (BLOCK (DECLARE RESULT 0)
             (DECLARE SUM 0)
             (FOR ((DECLARE I 0) (LOG< I 4) (+ I 1))
               (BLOCK (DECLARE ASGND (BITS A (+ (* 8 I) 7) (* 8 I)))
                      (DECLARE BSGND (BITS B (+ (* 8 I) 7) (* 8 I)))
                      (DECLARE SUMSGND (BITS (+ (SI ASGND 8) (SI BSGND 8)) 8 0))
                      (IF (LOG< (SI SUMSGND 9) -128)
                          (ASSIGN SUM (BITS -128 7 0))
                          (IF (LOG>= SUM 128)
                              (ASSIGN SUM (BITS 127 7 0))
                              (ASSIGN SUM (BITS (SI SUMSGND 9) 7 0))))
                      (ASSIGN RESULT (SETBITS RESULT 32 (+ (* 8 I) 7) (* 8 I) SUM))))
             (RETURN RESULT)))
\end{verbatim}
\end{footnotesize}
\caption{RAC parser output}\label{ast}
\end{figure}

Note also that variable types do not explicitly appear in the output.  The problem of translating a typed language to an untyped language is addressed
by the parser, mainly by converting implicit register evaluations and type conversions to explicit computations.  Thus, in the expression that is derived from the declaration
\begin{verbatim}
     si9 sumSgnd = aSgnd + bSgnd;
\end{verbatim}
the registers {\tt ASGND} and {\tt BSGND} are evaluated according to their type (computing their signed integer values), and when their
sum is assigned to the 9-bit register {\tt SUMSGND}, the low order 9 bits are extracted.  Evaluation and assignment of fixed-point registers are more
complicated, involving division and multiplication by appropriate powers of 2.

\section{The ACL2 Code Generator}\label{codegen}

The primary objective considered in the design of the code generator was simplicity, of both the program and its output.  Since the translation is
yet another component of the verification process that must be trusted, along with SLEC and ACL2, it should be as transparent as possible.  A
secondary consideration is executability: efficiency is not an overriding concern, but simulation is important for the purpose of testing.
A third objective is susceptibility of the output to formal analysis, but this was found to be in conflict with the first two and therefore ignored
in the design.  Instead, as discussed in Section~\ref{proof}, it is addressed by converting the object code to a more manageable and provably equivalent form.

\begin{figure}
\begin{footnotesize}
\begin{verbatim}
                 tuple<ui53, ui53, si13> normalize(ui11 expa, ui11 expb, ui52 mana, ui52 manb) {  
                   ui53 siga = mana, sigb = manb;                         
                   const uint bias = 0x3FF;                               
                   si13 expaShft, expbShft;                               
                   if (expa == 0) {                                       
                     ui6 clz = CLZ64(siga);                               
                     siga <<= clz;                                        
                     expaShft = 1 - clz;}                                 
                   else {                                                 
                     siga[52] = 1;                                        
                     expaShft = expa;}                                    
                   if (expb == 0) {                                       
                     ui6 clz = CLZ64(sigb);                               
                     sigb <<= clz;                                        
                     expbShft = 1 - clz;}                                 
                   else {                                                 
                     sigb[52] = 1;                                        
                     expbShft = expb;}                                    
                   si13 expQ = expaShft - expbShft + bias;                
                   return tuple<ui53, ui53, si13>(siga, sigb, expQ);}
		   

                 (DEFUND NORMALIZE (EXPA EXPB MANA MANB)
                   (LET ((SIGA MANA) (SIGB MANB) (BIAS 1023))
                     (MV-LET (SIGA EXPASHFT)
                             (IF1 (LOG= EXPA 0)
                                  (LET ((CLZ (CLZ64 SIGA)))
                                    (MV (BITS (ASH SIGA CLZ) 52 0)
                                        (BITS (- 1 CLZ) 12 0)))
                                  (MV (SETBITN SIGA 53 52 1) EXPA))
                       (MV-LET (SIGB EXPBSHFT)
                               (IF1 (LOG= EXPB 0)
                                    (LET ((CLZ (CLZ64 SIGB)))
                                      (MV (BITS (ASH SIGB CLZ) 52 0)
                                          (BITS (- 1 CLZ) 12 0)))
                                    (MV (SETBITN SIGB 53 52 1) EXPB))
                         (MV SIGA SIGB
                            (BITS (+ (- (SI EXPASHFT 13) (SI EXPBSHFT 13)) BIAS) 12 0))))))
\end{verbatim}
\end{footnotesize}
\caption{Normalization of the operands of a floating-point divider}\label{normalize}
\end{figure}

Obviously, there are various idiomatic differences between C and ACL2 to be managed in the translation.  Large constant arrays, which occur in RTL
designs to represent blocks of read-only memory, are converted to lists of values; variable arrays and {\tt struct}s are represented as alists.
Among the operators defined in {\tt lib/rac} are the array access and assignment functions {\tt AG} and {\tt AS}.

The difference between the boolean values native to C and ACL2 (1 and 0 vs. {\tt T} and {\tt NIL}) requires attention.  Along with the comparators
{\tt LOG<}, etc., that were mentioned in Section~\ref{parser}, {\tt lib/rac} includes a macro {\tt IF1}, which is similar to {\tt IF} but tests its
first argument against 0.

The primary difficulty faced in code generation, however, is the translation from an imperative to a functional paradigm.
Our overall strategy is based on the conversion of sequences of assignments to nested bindings, using {\tt LET}, {\tt LET*}, and {\tt MV-LET}.
Each statement in the function body except the last corresponds to one or more variables that are bound in this nest.  For each of these
statements, the translator generates a triple consisting of the following:
\begin{itemize}
\item [(1)] a list of the variables whose previous values are read by the statement;
\item [(2)] a list of the variables that are written by the statement;
\item [(3)] a term that evaluates to a multiple value consisting of the updated values of the variables of (2), or a single
value if (2) is a singleton.
\end{itemize}
The bindings of the nest are derived from these triples.  Each statement generates either a {\tt LET} or an {\tt MV-LET} depending on whether
(2) is a singleton.  Whenever possible, a nested sequence of {\tt LET}s is combined into a single {\tt LET} or {\tt LET*}.  The body of the nest is
generated from the final statement of the function body.

Statements that require {\tt MV-LET} include multiple-valued function calls and some conditional branching
statements.  In the latter case, the translation may be especially complicated, as in Figure~\ref{normalize}, which displays the translation of
a function that normalizes the operands of a floating-point divider and computes the biased exponent of the quotient.

Naturally, iteration is translated to recursion, with an auxiliary recursive function generated for every {\tt for} loop.
The arguments of this function consist of (a) the loop variable, (b) any previously assigned variables that are accessed inside the loop,
including those that occur in the loop initialization or test, and (c) the variables that are assigned within the loop and are not local
to the loop.  A multiple value is returned comprising the updated values of the variables of (c).  For example, for the recursive function
{\tt ADD8-LOOP-0} displayed in Figure~\ref{transadd8}, we have (a) {\tt I}, (b) {\tt A} and {\tt B}, and (c) {\tt SUM} and {\tt RESULT}.
Note that only one of the two values returned by the non-recursive call to this function is subsequently used.  For this reason, along
with various other possibilities for optimization that are ignored in the interest of simplifying the process, every translated RAC file
begins with these two lines to pacify ACL2:
\begin{verbatim}
     (SET-IGNORE-OK T)
     (SET-IRRELEVANT-FORMALS-OK T)
\end{verbatim}

\begin{figure}
\begin{footnotesize}
\begin{verbatim}
                    (DEFUN ADD8-LOOP-0 (I A B SUM RESULT)
                      (DECLARE (XARGS :MEASURE (NFIX (- 4 I))))
                      (IF (AND (INTEGERP I) (< I 4))
                          (LET* ((ASGND (BITS A (+ (* 8 I) 7) (* 8 I)))
                                 (BSGND (BITS B (+ (* 8 I) 7) (* 8 I)))
                                 (SUMSGND (BITS (+ (SI ASGND 8) (SI BSGND 8)) 8 0))
                                 (SUM (IF1 (LOG< (SI SUMSGND 9) -128)
                                           (BITS -128 7 0)
                                           (IF1 (LOG>= SUM 128)
                                                (BITS 127 7 0)
                                                (BITS (SI SUMSGND 9) 7 0))))
                                 (RESULT (SETBITS RESULT 32 
                                                  (+ (* 8 I) 7) (* 8 I) 
                                                  SUM)))
                            (ADD8-LOOP-0 (+ I 1) A B SUM RESULT))
                          (MV SUM RESULT)))

                    (DEFUN ADD8 (A B)
                      (LET ((RESULT 0) (SUM 0))
                        (MV-LET (SUM RESULT) (ADD8-LOOP-0 0 A B SUM RESULT)
                          RESULT)))
\end{verbatim}
\end{footnotesize}
\caption{Translation of the signed integer adder}\label{transadd8}
\end{figure}

The construction of this recursive function is similar to that of the top-level function, but the final statement of the body is not treated
specially.  Instead, the body of the nest of bindings is a recursive call in which the loop variable is replaced by its updated value.  The
resulting term becomes the left branch of an {\tt IF} expression, of which the right branch is simply the returned variable (if there is only
one) or a multiple value consisting of the returned variables (if there are more than one).  The test of the {\tt IF} is
\begin{alltt}
     (AND (INTEGERP {\it var}) (INTEGERP {\it limit}) {\it term})
\end{alltt}
where the test of the loop is {\it var op limit} and {\it term} is the result of converting the test to an expression that evaluates to {\tt T}
or {\tt NIL}.  (The second conjunct of this term is omitted when {\it limit} is a constant.)

\begin{figure}
\begin{footnotesize}
\begin{verbatim}
                       (DEFUN CLZ64-LOOP-0 (I N K C Z) ... )
                    
                       (DEFUN CLZ64-LOOP-1 (K N C Z)
                         (DECLARE (XARGS :MEASURE (NFIX (- 6 K))))
                         (IF (AND (INTEGERP K) (< K 6))
                             (LET ((N (FLOOR N 2)))
                               (MV-LET (C Z) (CLZ64-LOOP-0 0 N K C Z)
                                 (CLZ64-LOOP-1 (+ K 1) N C Z)))
                             (MV N C Z)))

                       (DEFUN CLZ64-LOOP-2 (I X Z C)
                         (DECLARE (XARGS :MEASURE (NFIX (- 64 I))))
                         (IF (AND (INTEGERP I) (< I 64))
                             (LET ((Z (AS I (LOGNOT1 (BITN X I)) Z))
                                   (C (AS I (BITS 0 5 0) C)))               
                               (CLZ64-LOOP-2 (+ I 1) X Z C))
                             (MV Z C)))

                       (DEFUN CLZ64 (X)
                         (LET ((ASSERT (IN-FUNCTION CLZ64 (LOG<> X 0)))
                               (Z NIL)
                               (C NIL))
                           (MV-LET (Z C) (CLZ64-LOOP-2 0 X Z C)
                             (LET ((N 64))
                               (MV-LET (N C Z) (CLZ64-LOOP-1 0 N C Z)
                                 (AG 0 C))))))
\end{verbatim}
\end{footnotesize}
\caption{Translation of the leading zero counter}\label{transclz}
\end{figure}

As shown in Figure~\ref{transclz}, the translation of the leading zero counter includes three auxiliary functions,
corresponding to the loop that initializes variables and the subsequent nested pair.
Note that this program contains an assertion, a type of statement that we have not discussed.  In RAC, as in C, an assertion has no semantic
import but is useful in testing and as documentation.  In this case, it indicates that {\tt CLZ64} is not intended to be applied
to the argument 0.  In the translation, it is represented by the binding of the dummy variable {\tt ASSERT} to a call to the macro
{\tt IN-FUNCTION}, defined as follows:
\begin{verbatim}
     (defmacro in-function (fn term)
       `(if1 ,term () (er hard ',fn "Assertion ~x0 failed" ',term)))
\end{verbatim}
Thus, attempted evaluation of {\tt (CLZ64 0)} results in a run-time error:
\begin{verbatim}
     HARD ACL2 ERROR in CLZ64:  Assertion (LOG<> X 0) failed
\end{verbatim}

\section{Proving Theorems about RAC Functions}\label{proof}

In this section, we address certain difficulties that arise in the course of proving ACL2 theorems about translated RAC functions.  The techniques
presented here have been employed in every Intel or Arm RAC-based verification effort.

Proving a theorem of interest pertaining to an RTL module of any complexity is best begun, according to our experience, by analyzing the design
and writing out an informal but detailed and rigorous proof.  Some properties of bit vectors are simple enough to be derived automatically with
{\tt gl}~\cite{centaur}, but even in these cases such a proof and the insight gained through its development are valuable.  More often, the written
proof must be checked by formalizing it step by step in a long sequence of ACL2 lemmas.

As an illustration, consider the function {\tt compare64} of Figure~\ref{compare64}, which compares the absolute values of two 64-bit signed integers.
The computation that it performs is more complicated than necessary, using a fast carry-save adder followed by a slower carry-propagate adder, but it is
also more efficient than the more obvious solutions.  This becomes clear upon examination of the following informal (and uncharacteristically chatty)
proof of correctness:\medskip

\begin{figure}
\begin{footnotesize}
\begin{verbatim}
                               bool compare64(ui64 a, ui64 b) {  
                                 bool sgnA = a[63], sgnB = b[63];
                                 bool cin = sgnA || !sgnB;
                                 ui64 sum = ~a ^ ~b;      
                                 ui64 carry = ((~a & ~b) << 1) | 1;
                                 ui64 add1, add2;
                                 if (sgnA && !sgnB) {
                                   add1 = sum;
                                   add2 = carry;}
                                 else {
                                   add1 = sgnA ? ui64(~a) : a;
                                   add2 = sgnB ? b : ui64(~b);}
                                 ui65 diff = add1 + add2 + cin;
                                 return !diff[64];}
\end{verbatim}
\end{footnotesize}
\caption{Signed integer comparison}\label{compare64}
\end{figure}

\noindent
{\bf Lemma} {\it Let $A$ and $B$ be the signed integers represented by 64-bit vectors $a$ and $b$.  Then}
\[
\mathit{compare64}(a, b) = 1 \Leftrightarrow |B| > |A|.
\]

{\sc  Proof}:  We shall examine the case $A<0$, $B \geq 0$; the other cases are simpler.  In this case, the function computes the complements
of the operands, the values of which are
\[
\verb!~!a[63:0] = 2^{64} - a - 1 = 2^{64} - (2^{64} - |A|) - 1 = |A| - 1
\]
and
\[
\verb!~!b[63:0] = 2^{64} - b - 1 = 2^{64} - |B| - 1.
\]
If we were to compute the 65-bit sum
\[
\mathit{diff} = \verb!~!a[63:0] + \verb!~!b[63:0] + 2 = (|A| - 1) + (2^{64} - |B| - 1) + 2 = 2^{64} + |A| - |B|,
\]
then we could simply look at the most significant bit $\mathit{diff}[64]$ to see whether $|A| \geq |B|$.  But while introducing a single carry-in to
a carry-propagate adder does not affect the complexity of the operation, adding 2 instead of 1 requires a separate incrementer, consuming nearly as
much time as a second addition.  Therefore, the function proceeds by computing the carry-save vectors
\[
\mathit{add1} = \mathit{sum} = \verb!~!a[63:0] \;\verb!^!\; \verb!~!b[63:0]
\]
and
\[
\mathit{add2} = \mathit{carry} = 2(\verb!~!a[63:0] \;\verb!&!\; \verb!~!b[63:0]) + 1,
\]
the sum of which, according to Lemma 8.2 of~\cite{book}, is
\[
\verb!~!a[63:0] + \verb!~!b[63:0] + 1 = 2^{64} + |A| - |B| -1,
\]
and then executes a single full addition to compute
\[
\mathit{diff} = \mathit{add1} + \mathit{add2} + 1 = 2^{64} + |A| - |B|.
\]
Thus, $\mathit{compare64}(a, b) = 1 \Leftrightarrow \mathit{diff}[64] = 0 \Leftrightarrow |B| > |A|$.~$\blacksquare$\medskip

The purpose of this example is to illustrate the nature of proofs of this sort: we derive a sequence of intermediate results pertaining to
various local variables, each assertion depending on previous assertions, until we arrive at the desired final result.  Next, we would like to
formalize this argument in a proof of the following, which refers to the translation displayed in Figure~\ref{transcompare64}:
\begin{verbatim}
     (defthm correctness-of-compare64
       (implies (and (bvecp a 64) (bvecp b 64))
                (equal (compare64 a b)
                       (if (> (abs (si b 64))
                              (abs (si a 64)))
                           1 0))))
\end{verbatim}

\begin{figure}
\begin{footnotesize}
\begin{verbatim}
                    (DEFUN COMPARE64 (A B)                    
                      (LET* ((SGNA (BITN A 63))
                             (SGNB (BITN B 63))
                             (CIN (LOGIOR1 SGNA (LOGNOT1 SGNB)))
                             (SUM (LOGXOR (BITS (LOGNOT A) 63 0)
                                          (BITS (LOGNOT B) 63 0)))
                             (CARRY (BITS (LOGIOR (ASH (LOGAND (BITS (LOGNOT A) 63 0)
                                                             (BITS (LOGNOT B) 63 0))
                                                       1)
                                                  1)
                                          63 0)))
                        (MV-LET (ADD1 ADD2)
                                (IF1 (LOGAND1 SGNA (LOGNOT1 SGNB))
                                     (MV SUM CARRY)
                                     (MV (BITS (IF1 SGNA (LOGNOT A) A) 63 0)
                                         (BITS (IF1 SGNB B (LOGNOT B)) 63 0)))
                                (LET ((DIFF (BITS (+ (+ ADD1 ADD2) CIN) 64 0)))
                                  (LOGNOT1 (BITN DIFF 64))))))
\end{verbatim}
\end{footnotesize}
\caption{Translation of {\tt compare64}}\label{transcompare64}
\end{figure}
Although this result is certainly simple enough for {\tt gl}, we shall use it as an illustration of the process of formalizing a mathematical proof
pertaining to a RAC program.  Even in this case, in which the sequence of assignments is considerably shorter than one derived from a typical RTL
module, we do not get far in this exercise before realizing that there is no convenient way to formalize the argument given above.  How can we prove
a lemma characterizing the bindings of the local variables {\tt ADD1} and {\tt ADD2} (or even state such a lemma) and then use it to prove another
lemma about the value of {\tt DIFF}?  We have developed a four-step procedure that solves this problem, as illustrated below:

\begin{itemize}

\item [(1)] {\it Introduce encapsulated constant functions corresponding to the arguments of the function of interest that are constrained to satisfy
the hypotheses of the conjectured statement of correctness:}

\begin{verbatim}
     (defund inputsp (a b)
       (and (bvecp a 64) (bvecp b 64)))

     (encapsulate (((a) => *) ((b) => *))
       (local (defun a () 0))
       (local (defun b () 0))
       (defthm inputs-ok (inputsp (a) (b))
         :hints (("Goal" :in-theory (enable inputsp)))))
\end{verbatim}

\begin{figure}
\begin{footnotesize}
\begin{verbatim}
      RTL !>(include-book "~/acl2/books/projects/rac/lisp/internal-fns-gen")
      RTL !>(const-fns-gen 'compare64 'r state)

      (DEFUNDD SGNA NIL (BITN (A) 63))

      (DEFUNDD SGNB NIL (BITN (B) 63))

      (DEFUNDD CIN NIL (LOGIOR1 (SGNA) (LOGNOT1 (SGNB))))

      (DEFUNDD SUM NIL
        (LOGXOR (BITS (LOGNOT (A)) 63 0)
                (BITS (LOGNOT (B)) 63 0)))
  
      (DEFUNDD CARRY NIL
        (BITS (LOGIOR (ASH (LOGAND (BITS (LOGNOT (A)) 63 0)
                                   (BITS (LOGNOT (B)) 63 0))
                           1)
                      1)
              63 0))
  
      (DEFUNDD ADD1 NIL
        (IF1 (LOGAND1 (SGNA) (LOGNOT1 (SGNB)))
             (SUM)
             (BITS (IF1 (SGNA) (LOGNOT (A)) (A)) 63 0)))
  
      (DEFUNDD ADD2 NIL
        (IF1 (LOGAND1 (SGNA) (LOGNOT1 (SGNB)))
             (CARRY)
             (BITS (IF1 (SGNB) (B) (LOGNOT (B))) 63 0)))

      (DEFUNDD DIFF NIL (BITS (+ (+ (ADD1) (ADD2)) (CIN)) 64 0))

      (DEFUNDD R NIL (LOGNOT1 (BITN (DIFF) 64)))

      (DEFTHMD COMPARE64-LEMMA
        (EQUAL (R) (COMPARE64 (A) (B)))
        :HINTS (("Goal" :DO-NOT '(PREPROCESS) :EXPAND :LAMBDAS
                        :IN-THEORY '(C SGNA SGNB CIN SUM CARRY ADD1 ADD2 DIFF COMPARE64)))))
\end{verbatim}
\end{footnotesize}
\caption{Automatically generated definitions}\label{cuong}
\end{figure}

\item [(2)] {\it Derive definitions of constant functions corresponding to the local variables directly from the variables' bindings, and prove that
the function maps the input constants to the output constants.}  This is a straightforward process, performed automatically by a tool developed by
Cuong Chau, which resides in {\tt books/projects/rac/lisp/}.  (It was previously done by hand, which was tedious, time-consuming, error-prone.)  The
output of the tool for the function {\tt compare64} is displayed in Figure~\ref{cuong}.  Note that the arguments of Chau's function {\tt const-fns-gen}
include the function to be transformed and names to be associated with the outputs.  The generated functions allow us to reason about the variables and
derive a sequence of lemmas that mirrors the informal proof outlined above.  Note that the theorem, generated with appropriate hints, is established by
the prover simply by expanding all relevant definitions.  This can result in unmanageable code explosion if a function definition is too long. With this
in mind, large RAC models must be designed with sufficient modularity to avoid this result.

\item [(3)] {\it Derive the required properties of the outputs:}

\begin{verbatim}
     (defthmd compare64-main
       (equal (r) (if (> (abs (si (b) 64)) (abs (si (a) 64))) 1 0)))
\end{verbatim}
Virtually all of the work lies in developing a proof script culminating in this theorem, i.e., a formalization of the informal argument presented
earlier.

\item[(4)] {\it Lift this result by functional instantiation to establish the original statement of correctness.}  The lemma to be instantiated
is a simple combination of the results of (2) amd (3):

\begin{verbatim}
     (defthmd lemma-to-be-lifted
       (equal (compare64 (a) (b))
              (if (> (abs (si (b) 64)) (abs (si (a) 64))) 1 0))
       :hints (("Goal" :use (compare64-lemma compare64-main))))
\end{verbatim}

A lemma is functionally instantiated by replacing some subset of the functions that appear in it with another set of functions.  The
prover is then obligated to prove the corresponding instances of all axioms that pertain to the first set, which are usually just their definitions.
In this case the only such axiom is the formula that is exported from the encapsulation that introduced the constants {\tt (a)} and {\tt (b)}.
The {\tt :use} hint is formulated as follows:

\begin{Verbatim}[samepage=true]
   (defthm correctness-of-compare64
     (implies (inputsp a b)
              (equal (compare64 a b)
                     (if (> (abs (si b 64)) (abs (si a 64)))
                         1 0)))
     :hints (("Goal" :use (:functional-instance lemma-to-be-lifted
                            (a (lambda () (if (inputsp a b) a (a))))
                            (b (lambda () (if (inputsp a b) b (b))))))))
\end{Verbatim}
The choice of instantiation is perhaps best understood by examining the subgoals that it generates.  {\tt Subgoal 2} is just the top-level goal
with the indicated functional instance inserted as a hypothesis; {\tt Subgoal 1} is the incurred proof obligation mentioned above:

\begin{Verbatim}[samepage=true]
     ... We are left with the following two subgoals.

     Subgoal 2
     (IMPLIES (EQUAL (COMPARE64 (IF (INPUTSP A B) A (A))
                                (IF (INPUTSP A B) B (B)))
                     (IF (< (ABS (SI (IF (INPUTSP A B) A (A)) 64))
                            (ABS (SI (IF (INPUTSP A B) B (B)) 64)))
                         1 0))
              (IMPLIES (INPUTSP A B)
                       (EQUAL (COMPARE64 A B)
                              (IF (< (ABS (SI A 64)) (ABS (SI B 64)))
                                  1 0)))).

     But we reduce the conjecture to T, by case analysis.

     Subgoal 1
     (INPUTSP (IF (INPUTSP A B) A (A))
              (IF (INPUTSP A B) B (B))).

     This simplifies, using the simple :rewrite rule INPUTS-OK, to T.

Q.E.D.
\end{Verbatim}

Finally, we note that the book containing {\tt const-fns-gen} includes a second program, {\tt loop-fns\-gen}, which is critical in dealing with
iterative designs, dividers in particular.  Such a module is typically modeled by a large {\tt for} loop with a loop variable that is incremented
on each iteration.  In this situation, it is necessary to reason about the value of a variable that is computed in a given iteration and to relate
it to values computed in preceding interations.  Thus, instead of constant functions, {\tt loop-fns-gen} generates a function of a single
argument, representing the loop variable, for each variable that is updated iteratively.  Space does not allow an illustration of this technique
here, but examples may be found in the {\tt fdiv} and {\tt fsqrt} subdirectories of {\tt books/projects/arm}.

\end{itemize}

\nocite{*}
\bibliographystyle{eptcs}
\bibliography{rac}
\end{document}